\documentclass[english,aps,tightenlines,showpacs, showkeys, notitlepage]{revtex4-1}
\usepackage[latin9]{inputenc}
\usepackage{amsmath}
\usepackage{amssymb}

\makeatletter
\usepackage{dcolumn}
\usepackage{bm}

\makeatother

\usepackage{babel}
\begin{document}
\title{$D=4$ supergravity from the Maxwell-Weyl superalgebra}
\author{Salih Kibaro\u{g}lu$^{1,2}$}
\email{salihkibaroglu@gmail.com}

\author{Oktay Cebecio\u{g}lu$^{1}$}
\email{ocebecioglu@kocaeli.edu.tr}

\date{\today}
\begin{abstract}
We present the construction of the first-order $D=4$, $\mathcal{N}=1$
supergravity action by gauging the Maxwell-Weyl superalgebra. The
four-form lagrangian is constructed by using the curvatures of the
algebra and the local scale invariance of the action is achieved through
the introduction of a compensating scalar field. Finally, we find
the generalized Einstein equation with a coordinate dependent cosmological
term. 
\end{abstract}
\affiliation{$^{1}$Department of Physics, Kocaeli University, 41380 Kocaeli, Turkey}
\affiliation{$^{2}$Institute for Theoretical Physics, University of Wroc\l aw,
Pl. Maksa Borna 9, Pl\textendash 50-204 Wroc\l aw, Poland}
\maketitle

\section{Introduction}

The Maxwell symmetry appears if the Minkowski space-time is filled
with an additional background field \citep{bacry1970,schrader1972}.
In other words, this symmetry can be interpreted as a modification
of the Poincaré symmetry which describes the empty Minkowski space-time,
and the translation generators are no longer abelian but satisfy \citep{soroka2005},
\begin{equation}
\left[P_{a},P_{b}\right]=iZ_{ab},
\end{equation}
where the six additional antisymmetric generators $Z_{ab}$ transform
as a second rank tensor under the action of Lorentz group. In early
studies, this background field was associated with constant electromagnetic
(EM) fields. Nowadays interpretation of the background field as well
as this new algebra have opened up new directions. For example, this
algebra was extensively studied to generalize Einstein's theory of
gravity. In \citep{azcarraga2011,soroka2012,durka2011,cebecioglu2014},
the generalized cosmological constant and additional interaction terms
were derived alternatively by extending the theory of gravity based
on the Maxwell algebra. This fact may have a fundamental importance
since the studies and observations on the cosmological constant, the
dark energy and the cosmic microwave background indicate that there
should be a background field filling our space-time. We also know
that there is a close connection between cosmological constant and
dark energy \citep{frieman2008,padmanabhan2009}, so the Maxwell symmetries
may provide a powerful geometrical framework for these significant
topics. 

The supersymmetric extension of the Maxwell algebra was presented
in \citep{bonanos2010} and it describes the geometry of $D=4$, $\mathcal{N}=1$
superspace with a constant abelian supersymmetric gauge field background.
This modification of superspace is known as the superMaxwell space.
Contrary to the conventional superalgebras, this superalgebra contains
two Majorana spinor generators $Q_{\alpha}$ and $\Sigma_{\alpha}$.
This superalgebra can be considered as a generalization of the D\textquoteright Auria-Fré
superalgebra \citep{dAuria1982} and the Green algebra \citep{green1989}
which have additional fermionic generators. Besides, the Maxwell superalgebras
were also obtained in \citep{azcarraga2013,concha2014A} by using
algebraic expansion methods \citep{azcarraga2004,izaurieta2006}.
In \citep{azcarraga2014}, the authors derived the first order $D=4$,
$\mathcal{N}=1$ pure supergravity lagrangian four-form by using the
curvatures of the Maxwell superalgebra. Subsequently, the generalized
supersymmetric cosmological constant was constructed based on the
Maxwell superalgebras in $\mathcal{N}=1$ case \citep{concha2015A,durka2012}.
Beyond the case of $\mathcal{N}=1$, the $\mathcal{N}-$extended Maxwell
superalgebras were considered in \citep{azcarraga2013,concha2014A,lukierski2011,kamimura2012}.
Recent developments and some interesting studies about the Maxwell
(super)algebras can be found in \citep{gomis2009,bonanos2009,bonanos2010A,fedoruk2013,fedoruk2013A,fedoruk2012,hoseinzadeh2014,concha2014,concha2015B,cebecioglu2015,kibaroglu2018,penafiel2017,penafiel2018,ravera2018}.

The Weyl enlargement of the Maxwell algebra, which named as the Maxwell-Weyl
algebra $\mathcal{MW}$, and its supersymmetric extension $s\mathcal{MW}$
were firstly presented in \citep{bonanos2010A}. In our earlier work,
we constructed a gauge theory of gravity based on $\mathcal{MW}$
and obtained the generalized Einstein equation with cosmological constant
\citep{cebecioglu2014}. The main purpose of this letter is to establish
a gauge theory of gravity based on $s\mathcal{MW}$ and to construct
the first-order $D=4$, $\mathcal{N}=1$ supergravity action. 

The paper organized as follows. In Section 2, we give a brief summary
of the $\mathcal{MW}$ algebra and its gravity action. In Section
3, we introduce $s\mathcal{MW}$ algebra and obtain transformation
of the gauge fields, the curvatures and the Bianchi identities of
the algebra. In Section 4, the supergravity action is constructed
by using the curvatures of $s\mathcal{MW}$ algebra together with
an additional compensating scalar field. In Section 5, we conclude
the paper with some comments. In the last section, our notations and
conventions are given.

\section{Maxwell-Weyl algebra and the gravity action}

In this section, we briefly give the results of our previous study
\citep{cebecioglu2014}. The Maxwell-Weyl algebra $\mathcal{MW}$
can be considered as Weyl enlargement of the Maxwell algebra with
a dilatation generator \citep{bonanos2010A}. The non-zero commutation
relations of $\mathcal{MW}$ algebra are given by,

\begin{eqnarray}
\left[M_{ab},M_{cd}\right] & = & i\left(\eta_{ad}M_{bc}+\eta_{bc}M_{ad}-\eta_{ac}M_{bd}-\eta_{bd}M_{ac}\right),\nonumber \\
\left[P_{a},P_{b}\right] & = & iZ_{ab},\nonumber \\
\left[M_{ab},P_{c}\right] & = & i\left(\eta_{bc}P_{a}-\eta_{ac}P_{b}\right),\nonumber \\
\left[M_{ab},Z_{cd}\right] & = & i\left(\eta_{ad}Z_{bc}+\eta_{bc}Z_{ad}-\eta_{ac}Z_{bd}-\eta_{bd}Z_{ac}\right),\nonumber \\
\left[P_{a},D\right] & = & iP_{a},\nonumber \\
\left[Z_{ab},D\right] & = & 2iZ_{ab},\label{eq: mw algebra}
\end{eqnarray}
where $\eta_{ab}$ is the Minkowski metric which has $diag\left(\eta_{ab}\right)=\left(+,-,-,-\right)$
and the indices $a,b,...=0,...,3$. In addition to the Weyl algebra,
this algebra contains six new additional tensorial generators $Z_{ab}$. 

Let us start to construct the gauge theory of gravity based on $\mathcal{MW}$
algebra. We first introduce the following one-form gauge field with,

\begin{equation}
\mathcal{A}=e^{a}P_{a}+B^{ab}Z_{ab}+\chi D-\frac{1}{2}\omega^{ab}M_{ab},\label{eq: gauge field 1}
\end{equation}
where the coefficient fields $e^{a}\left(x\right)$, $B^{ab}\left(x\right)$,
$\chi\left(x\right)$ and $\omega^{ab}\left(x\right)$ are the one-form
gauge fields for the corresponding generators, respectively. Using
the structure equation $\mathcal{F}=d\mathcal{A}+\frac{i}{2}\left[\mathcal{A},\mathcal{A}\right]$
and defining the curvatures as $\mathcal{F}=F^{a}P_{a}+F^{ab}Z_{ab}+fD-\frac{1}{2}R^{ab}M_{ab}$,
we find the associated two-form curvatures as, 

\begin{eqnarray}
F^{a} & = & de^{a}+\omega_{\,\,\,b}^{a}\wedge e^{b}+\chi\wedge e^{a},\nonumber \\
F^{ab} & = & dB^{ab}+\omega_{\,\,\,\,c}^{[a}\wedge B^{c|b]}+2\chi\wedge B^{ab}-\frac{1}{2}e^{a}\wedge e^{b},\nonumber \\
f & = & d\chi,\nonumber \\
R^{ab} & = & d\omega^{ab}+\omega_{\,\,\,c}^{a}\wedge\omega^{cb}.
\end{eqnarray}
The infinitesimal gauge transformation of the curvatures can be found
by introducing the Lie algebra valued parameters,

\begin{equation}
\zeta\left(x\right)=y^{a}\left(x\right)P_{a}+\varphi^{ab}\left(x\right)Z_{ab}+\rho\left(x\right)D-\frac{1}{2}\tau^{ab}\left(x\right)M_{ab},\label{eq: aux field mw}
\end{equation}
with the help of the equation $\delta\mathcal{F}=i\left[\zeta,\mathcal{F}\right]$.
Here, $y^{a}\left(x\right)$, $\varphi^{ab}\left(x\right)$, $\rho\left(x\right)$
and $\tau^{ab}\left(x\right)$ are the space-time translations, translation
in tensorial space, dilatation parameter, and the Lorentz transformation
parameters respectively. Thus, the transformations of the curvatures
under $\mathcal{MW}$ algebra are found as follows,

\begin{eqnarray}
\delta F^{a} & = & \omega_{\,\,b}^{a}F^{b}+\rho F^{a}-R_{\,\,b}^{a}y^{b}-fy^{a},\nonumber \\
\delta F^{ab} & = & \omega_{\,\,\,c}^{[a}F^{c|b]}+2\rho F^{ab}+\frac{1}{2}F^{[a}y^{b]}-R_{\,\,\,c}^{[a}\varphi^{c|b]}-\frac{1}{2}f\varphi^{ab},\nonumber \\
\delta f & = & 0,\nonumber \\
\delta R^{ab} & = & \omega_{\,\,\,c}^{[a}R^{c|b]}.
\end{eqnarray}
Taking the covariant derivative of given curvatures, the corresponding
Bianchi identities can be found as follows
\begin{eqnarray}
\mathcal{D}F^{a} & = & R_{\,\,\,b}^{a}\wedge e^{b}+f\wedge e^{a},\nonumber \\
\mathcal{D}F^{ab} & = & R_{\,\,\,\,c}^{[a}\wedge B^{c|b]}+2f\wedge B^{ab}-\frac{1}{2}F^{[a}\wedge e^{b]},\nonumber \\
\mathcal{D}f & = & 0,\nonumber \\
\mathcal{D}R^{ab} & = & 0,
\end{eqnarray}
where $\mathcal{D}\Phi=[d+\omega+w(\Phi)\chi]$$\Phi$ is the Lorentz-Weyl
covariant derivative and $w$ being the Weyl weight of the corresponding
field. 

Construction of a lagrangian based on $\mathcal{MW}$ algebra is slightly
different from the ordinary Maxwell algebra because $\mathcal{MW}$
algebra contains the scale transformations which act on fields as

\begin{equation}
\Phi'\left(x\right)=e^{w\rho\left(x\right)}\Phi\left(x\right)
\end{equation}
where $\rho\left(x\right)$ is local scale parameter and $w$ is the
Weyl weight of scalar field $\Phi\left(x\right)$ (for more detail
about the Weyl transformation see \citep{omote1971,bregman1973,charap1974,kasuya1975,blagojevic2002,babourova2006}).
In our context, transformation of the metric tensor with respect to
the Weyl transformation is $g'_{\mu\nu}=e^{2\rho\left(x\right)}g_{\mu\nu}$.
Thus, the variation of metric tensor satisfies 
\begin{equation}
\delta g_{\mu\nu}\left(x\right)=2\rho\left(x\right)g_{\mu\nu}\left(x\right)\label{eq: var metric}
\end{equation}
under infinitesimal gauge transformations, where $\mu,\nu=0,...,3$
are space-time indices. This means that the Weyl weight of the metric
tensor is $w\left(g_{\mu\nu}\right)=2$, $w\left(g^{\mu\nu}\right)=-2$
and $w\left(\sqrt{-g}\right)=4$. For this reason, the well-known
action $S=\int d^{4}x\sqrt{-g}R$ is not invariant under the local
scale transformations. To overcome this difficulty, Weyl used quadratic
terms in the action to construct a consistent theory \citep{weyl1961}.
Furthermore, Brans and Dicke constructed a lagrangian with the combination
of the scalar curvature $R$ and a new compensating scalar field $\phi\left(x\right)$
which have a Weyl weight \citep{brans1961}. Later, Dirac used the
scalar field $\phi\left(x\right)$ with having $w\left(\phi\right)=-1$
and constructed the action by using $\phi^{2}R$ rather than $R$
\citep{dirac1973} (for more information see \citep{dereli1982,agnese1975}). 

From these theoretical backgrounds, in accordance with Dirac's convention,
we introduced a compensating scalar field $\phi\left(x\right)$ which
transforms as $\delta\phi=-\rho\phi$ in order to make the Einstein-Hilbert
action to be local scale invariant (more detail see \citep{cebecioglu2014,dirac1973}).
Now, if we define a shifted curvature as $\mathcal{J}^{ab}=R^{ab}+2\gamma\phi^{2}F^{ab}$,
we can write the following action having Weyl weight zero,

\begin{eqnarray}
S & = & \int\frac{1}{2\kappa\gamma}\mathcal{J}\wedge^{\ast}\mathcal{J}+\frac{1}{4}f\wedge^{\ast}f-\frac{1}{2}\mathcal{D}\phi\wedge^{\ast}\mathcal{D}\phi+\frac{\lambda}{4}\phi^{4}{}^{\ast}1,
\end{eqnarray}
where $\kappa$, $\gamma$ and $\lambda$ are constants and $*$ represents
the Hodge duality operator.

\section{Gauging the maxwell-weyl superalgebra}

In this section, we consider gauge theory of the $s\mathcal{MW}$
algebra. The Weyl enlargement of the Maxwell superalgebra was first
proposed in \citep{bonanos2010A} with a different notation. In our
convention, we can write the $s\mathcal{MW}$ algebra in addition
to Eq.(\ref{eq: mw algebra}) as 

\begin{eqnarray}
\left\{ Q_{\alpha},Q_{\beta}\right\}  & = & 2\left(C\gamma^{c}\right)_{\alpha\beta}P_{c},\nonumber \\
\left[P_{a},Q_{\beta}\right] & = & -i\left(\gamma_{a}\Sigma\right)_{\beta},\nonumber \\
\left\{ Q_{\alpha},\Sigma_{\beta}\right\}  & = & -\frac{i}{2}\left(C\sigma^{cd}\right)_{\alpha\beta}Z_{cd}-i\left(C\gamma_{5}\right)_{\alpha\beta}B,\nonumber \\
\left[M_{ab},Q_{\rho}\right] & = & \frac{1}{2}\left(\sigma_{ab}Q\right)_{\rho},\,\,\,\,\,\,\,\,\,\left[M_{ab},\Sigma_{\rho}\right]=\frac{1}{2}\left(\sigma_{ab}\Sigma\right)_{\rho},\nonumber \\
\left[D,B\right] & = & -2iB,\nonumber \\
\left[D,Q_{\alpha}\right] & = & -\frac{i}{2}Q_{\alpha},\,\,\,\,\,\,\,\,\,\left[D,\Sigma_{\alpha}\right]=-\frac{3i}{2}\Sigma_{\alpha}.
\end{eqnarray}
Here $Q_{\alpha}$ are the standard supersymmetric fermionic charges
where the spinor indices $\alpha,\beta,...=1,...,4$. To close the
commutator $\left[P,Q\right]$ satisfying at the same time the Jacobi
identities $\left(P,Q,Q\right)$, we need to introduce an additional
fermionic generator $\Sigma_{\alpha}$ as in \citep{bonanos2010,bonanos2010A}.
In first glance, the presence of an extra fermionic charge in the
theory means a second gravitino, but it is not so. As we will see
later, the sigma generator does not contribute the spacetime coordinates.
Only contribution from the fermionic charges to the spacetime coordinates
comes from $Q_{\alpha}$ supercharges. In $\mathcal{N}=1$ case, the
motivation for introducing extra fermionic generators can be found
in \citep{dAuria1982,green1989} and they can also be obtained by
taking into account of the algebraic expansion methods \citep{azcarraga2004,izaurieta2006}.
The central charge $B$ describes the off-shell extension of $U\left(1\right)$
field strength multiplet in $D=4$. If we remove $B$, which is mathematically
possible, we get the minimal Maxwell-Weyl superalgebra (for more detail
see \citep{bonanos2010A}). For the closure of this superalgebra,
we need to use the cyclic identities given as $\left(C\gamma_{e}\right)_{(\alpha\beta}\left(C\gamma^{e}\right)_{\rho\sigma)}=0$.
All spinors given in this paper are characterized by Majorana spinors
and satisfying the generic relation $\bar{Q}=Q^{T}C$ where $C$ is
the charge conjugation matrix. 

To gauge this algebra, we start by introducing a one-form connection
$\mathcal{A}\left(x\right)$ as,

\begin{equation}
\mathcal{A}\left(x\right)=e^{a}P_{a}+B^{ab}Z_{ab}+\chi D-\frac{1}{2}\omega^{ab}M_{ab}+\psi^{\alpha}Q_{\alpha}+\xi^{\alpha}\Sigma_{\alpha}+AB.\label{eq:gauge_field 2}
\end{equation}
Comparing with Eq.(\ref{eq: gauge field 1}), the last expression
contains three additional gauge fields $\psi^{\alpha}\left(x\right)$,
$\xi^{\alpha}$$\left(x\right)$, and $A\left(x\right)$ which correspond
to the gravitino field, the additional fermionic gauge field and gauge
field of central charge $B$, respectively. The variations of these
fields under infinitesimal gauge transformation of $s\mathcal{MW}$
can be obtained by $\delta\mathcal{A}=-d\zeta+i\left[\zeta,\mathcal{A}\right]$,
where $\zeta\left(x\right)$ is the Lie algebra valued auxiliary field
defined as,

\begin{equation}
\zeta\left(x\right)=y^{a}P_{a}+\varphi^{ab}Z_{ab}+\rho D-\frac{1}{2}\tau^{ab}M_{ab}+\varepsilon^{\alpha}Q_{\alpha}+v^{\alpha}\Sigma_{\alpha}+rB.\label{eq:Auxiliary_field}
\end{equation}
In addition to Eq.(\ref{eq: aux field mw}), $\varepsilon^{\alpha}\left(x\right)$,
$v^{\alpha}\left(x\right)$ and $r\left(x\right)$ represent the parameters
of $Q_{\alpha}$, $\Sigma_{\alpha}$ and $B$, respectively. From
these definitions, one can find transformation of the gauge fields
as,

\begin{eqnarray}
\delta e^{a} & = & -\mathcal{\mathfrak{\mathcal{D}}}y^{a}+\tau_{\,\,\,b}^{a}e^{b}+\rho e^{a}-2i\bar{\varepsilon}\gamma^{a}\psi,\nonumber \\
\delta B^{ab} & = & -\mathcal{\mathfrak{\mathcal{D}}}\varphi^{ab}+\tau_{\,\,\,\,c}^{[a}B^{c|b]}+2\rho B^{ab}+\frac{1}{2}e^{[a}y^{b]}\nonumber \\
 &  & -\frac{1}{2}\bar{\varepsilon}\sigma^{ab}\xi-\frac{1}{2}\bar{v}\sigma^{ab}\psi,\nonumber \\
\delta\chi & = & -d\rho,\nonumber \\
\delta\omega^{ab} & = & -\mathcal{\mathfrak{\mathcal{D}}}\tau^{ab},\nonumber \\
\delta\psi^{\alpha} & = & -\mathcal{\mathfrak{\mathcal{D}}}\varepsilon^{\alpha}-\frac{i}{4}\tau^{ab}\left(\sigma_{ab}\psi\right)^{\alpha}+\frac{1}{2}\rho\psi^{\alpha},\nonumber \\
\delta\xi^{\alpha} & = & -\mathcal{\mathfrak{\mathcal{D}}}v^{\alpha}-\frac{i}{4}\tau^{ab}\left(\sigma_{ab}\xi\right)^{\alpha}+\frac{3}{2}\rho\xi^{\alpha}\nonumber \\
 &  & +y^{c}\left(\gamma_{c}\psi\right)^{\alpha}-e^{c}\left(\gamma_{c}\varepsilon\right)^{\alpha},\nonumber \\
\delta A & = & -\mathcal{\mathfrak{\mathcal{D}}}r-2\rho A-\bar{\varepsilon}\gamma_{5}\xi+\bar{v}\gamma_{5}\psi,\label{eq: var gauge f}
\end{eqnarray}
where $\mathcal{D}$ is the Lorentz-Weyl covariant derivative which
was given in Section 2. To find the associated two-form curvatures
of corresponding superalgebra, we use the well-known relation $\mathcal{F}=d\mathcal{A}+\frac{i}{2}\left[\mathcal{A},\mathcal{A}\right]$,
where $\mathcal{F}$ represents the curvatures and it is given in
the following form,

\begin{equation}
\mathcal{F}=F^{a}P_{a}+F^{ab}Z_{ab}+fD-\frac{1}{2}R^{ab}M_{ab}+\Psi^{\alpha}Q_{\alpha}+\Xi^{\alpha}\Sigma_{\alpha}+KB.\label{eq:curvature}
\end{equation}
Therefore, the explicit form of the curvatures can be found as follows,

\begin{eqnarray}
F^{a} & = & de^{a}+\omega_{\,\,\,b}^{a}\wedge e^{b}+\chi\wedge e^{a}-i\left(\bar{\psi}\wedge\gamma^{a}\psi\right),\nonumber \\
F^{ab} & = & dB^{ab}+\omega_{\,\,\,\,c}^{[a}\wedge B^{c|b]}+2\chi\wedge B^{ab}\nonumber \\
 &  & -\frac{1}{2}e^{a}\wedge e^{b}-\frac{1}{2}\left(\bar{\psi}\wedge\sigma^{ab}\xi\right),\\
f & = & d\chi,\nonumber \\
R^{ab} & = & d\omega^{ab}+\omega_{\,\,\,c}^{a}\wedge\omega^{cb},\nonumber \\
\Psi^{\alpha} & = & d\psi^{\alpha}-\frac{i}{4}\omega^{ab}\wedge\left(\sigma_{ab}\psi\right)^{\alpha}+\frac{1}{2}\chi\wedge\psi^{\alpha},\nonumber \\
\varXi^{\alpha} & = & d\xi^{\alpha}-\frac{i}{4}\omega^{ab}\wedge\left(\sigma_{ab}\xi\right)^{\alpha}+\frac{3}{2}\chi\wedge\xi^{\alpha}+e^{c}\wedge\left(\gamma_{c}\psi\right)^{\alpha},\nonumber \\
K & = & dA+2\chi\wedge A-\left(\bar{\psi}\gamma_{5}\xi\right).\label{eq: curvatures smw}
\end{eqnarray}
The curvatures have the following length dimensions: $\left[F^{a}\right]=L$,
$\left[F^{ab}\right]=L^{2}$, $\left[f\right]=L^{0}$, $\left[R^{ab}\right]=L^{0}$,
$\left[\Psi^{\alpha}\right]=L^{1/2}$, $\left[\Xi^{\alpha}\right]=L^{3/2}$
and $\left[K\right]=L^{2}$. Here, if we take $\mathcal{F}=0$, we
find the Maurer-Cartan equations for $s\mathcal{MW}$. The infinitesimal
gauge transformation of the curvatures under $s\mathcal{MW}$ group
can be found by using the relation $\delta\mathcal{F}=i\left[\zeta,\mathcal{F}\right]$
as,

\begin{eqnarray}
\delta F^{a} & = & \omega_{\,\,\,b}^{a}F^{b}+\rho F^{a}-R_{\,\,\,b}^{a}y^{b}-fy^{a}-2i\bar{\varepsilon}\gamma^{a}\Psi,\nonumber \\
\delta F^{ab} & = & \omega_{\,\,\,\,c}^{[a}F^{c|b]}+2\rho F^{ab}+\frac{1}{2}F^{[a}y^{b]}-R_{\,\,\,\,c}^{[a}\varphi^{c|b]}-\frac{1}{2}f\varphi^{ab}\nonumber \\
 &  & -\frac{1}{2}\bar{\varepsilon}\sigma^{ab}\Xi-\frac{1}{2}\bar{v}\sigma^{ab}\Psi,\nonumber \\
\delta f & = & 0,\nonumber \\
\delta R^{ab} & = & \omega_{\,\,\,\,c}^{[a}R^{c|b]},\nonumber \\
\delta\Psi^{\alpha} & = & -\frac{i}{4}\tau^{ab}\left(\sigma_{ab}\Psi\right)^{\alpha}+\frac{1}{2}\rho\Psi^{\alpha}+\frac{i}{4}R^{ab}\left(\sigma_{ab}\varepsilon\right)^{\alpha}-\frac{1}{2}\varepsilon^{\alpha}f,\nonumber \\
\delta\Xi^{\alpha} & = & -\frac{i}{4}\tau^{ab}\left(\sigma_{ab}\Xi\right)^{\alpha}+\frac{3}{2}\rho\Xi^{\alpha}+\frac{i}{4}R^{ab}\left(\sigma_{ab}v\right)^{\alpha}\nonumber \\
 &  & -\frac{3}{2}v^{\alpha}f+y^{c}\left(\gamma_{c}\Psi\right)^{\alpha}+F^{c}\left(\gamma_{c}\varepsilon\right)^{\alpha},\nonumber \\
\delta K & = & 2\rho K-2rf-\bar{k}\gamma_{5}\Xi+\bar{v}\gamma_{5}\Psi.
\end{eqnarray}
Finally, the Bianchi identities of corresponding curvatures can be
obtained as follows,

\begin{eqnarray}
\mathcal{\mathcal{D}}F^{a} & = & R_{\,\,\,b}^{a}\wedge e^{b}+f\wedge e^{a}+2i\bar{\psi}\gamma^{a}\Psi,\nonumber \\
\mathcal{\mathcal{D}}F^{ab} & = & R_{\,\,\,\,c}^{[a}\wedge B^{c|b]}+2f\wedge B^{ab}-\frac{1}{2}F^{[a}\wedge e^{b]}-\frac{1}{2}\bar{\Psi}\wedge\sigma^{ab}\xi\nonumber \\
 &  & +\frac{1}{2}\bar{\psi}\wedge\sigma^{ab}\Xi+\frac{1}{2}e^{c}\wedge\left(\bar{\psi}\wedge\sigma^{ab}\gamma_{c}\psi\right),\nonumber \\
\mathcal{\mathcal{D}}f & = & 0,\nonumber \\
\mathcal{\mathcal{\mathcal{D}}}R^{ab} & = & 0,\nonumber \\
\mathcal{\mathfrak{\mathcal{\mathcal{D}}}}\Psi^{\alpha} & = & -\frac{i}{4}R^{ab}\wedge\left(\sigma_{ab}\psi\right)^{\alpha}+\frac{1}{2}f\wedge\psi^{\alpha},\nonumber \\
\mathcal{\mathcal{\mathcal{D}}}\Xi^{\alpha} & = & -\frac{i}{4}R^{ab}\wedge\left(\sigma_{ab}\xi\right)^{\alpha}+\frac{3}{2}f\wedge\xi^{\alpha}+F^{c}\wedge\left(\gamma_{c}\psi\right)^{\alpha}-e^{c}\wedge\left(\gamma_{c}\Psi\right)^{\alpha},\nonumber \\
\mathcal{\mathfrak{\mathcal{\mathcal{D}}}}K & = & 2f\wedge A-\bar{\Psi}\wedge\gamma_{5}\xi+\bar{\psi}\wedge\gamma_{5}\Xi.\label{eq: bianchi}
\end{eqnarray}

\section{Constructing $D=4$, $\mathcal{N}=1$ supergravity action}

To construct scale invariant theory of gravity, as we mentioned in
Section 2, we will use the method of the Brans-Dicke theory of gravitation
\citep{brans1961} with the Dirac formalism \citep{dirac1973} which
contains a scalar compensating field $\phi\left(x\right)$ with $w\left(\phi\right)=-1$.
Following a similar method to \citep{cebecioglu2014,azcarraga2014,fradkin1985},
we will establish a lagrangian by using the curvature bilinear constructed
out of $s\mathcal{MW}$ algebra together with the additional scalar
field (more details on supergravity including the local scale transformation
can be found in \citep{kaku1978,nieuwenhuizen1986,nieuwenhuizen1981}). 

In $s\mathcal{MW}$ framework, the metric tensor shows the same characteristics
with Eq.(\ref{eq: var metric}) so it has $w\left(g_{\mu\nu}\right)=2$.
According to Eq.(\ref{eq: var gauge f}), the Weyl weights of the
gauge fields are defined as $w\left(e_{\mu}^{a}\right)=1,$ $w\left(B_{\mu}^{ab}\right)=2,$
$w\left(\chi_{\mu}\right)=0,$ $w\left(\omega_{\mu}^{ab}\right)=0,$
$w\left(\psi_{\mu}^{\alpha}\right)=1/2,$ $w\left(\xi_{\mu}^{\alpha}\right)=3/2,$
and $w\left(A_{\mu}\right)=2$. Similarly, the Weyl weights of the
curvatures can be found from Eq.(\ref{eq: curvatures smw}). Finally,
we can write a free gravitational four-form lagrangian density as
follows,

\begin{equation}
L_{f}=\frac{\phi^{2}}{2\kappa}\left(\varepsilon_{abcd}R^{ab}\wedge F^{cd}+4\bar{\Xi}\wedge\gamma_{5}\Psi-2f\wedge K\right),
\end{equation}
where $\kappa$ is a constant. In addition, one can use the following
lagrangian for vacuum,

\begin{equation}
L_{0}=\frac{1}{4}f\wedge^{*}f-\frac{1}{2}\mathfrak{\mathcal{\mathcal{D}}}\phi\wedge^{*}\mathcal{\mathcal{D}}\phi+\frac{\lambda}{4}\phi^{4*}1,
\end{equation}
where $\lambda$ is another constant, the first term represents a
Maxwell like kinetic term, the second corresponds to the kinetic term
for the compensating scalar field and the last one is a self-interaction
term for the scalar field. Finally, the total action can be written
as follows,

\begin{equation}
S=\int L_{f}+L_{0}.\label{eq:lagrangian_euler}
\end{equation}
In order to simplify the total action, let us expand the free part
$L_{f}$ as follows,

\begin{eqnarray}
S_{f} & = & \frac{1}{2\kappa}\int\phi^{2}\varepsilon_{abcd}R^{ab}\wedge\mathcal{\mathcal{D}}B^{cd}-\frac{1}{2}\phi^{2}\varepsilon_{abcd}R^{ab}\wedge e^{c}\wedge e^{d}\nonumber \\
 &  & -\frac{1}{2}\phi^{2}\varepsilon_{abcd}R^{ab}\wedge\left(\bar{\xi}\wedge\sigma^{cd}\psi\right)+4\phi^{2}\mathcal{\mathfrak{\mathcal{\mathcal{D}}}}\bar{\xi}\wedge\gamma_{5}\Psi\nonumber \\
 &  & +4\phi^{2}\bar{\psi}\wedge e_{a}\gamma^{a}\gamma_{5}\wedge\Psi-2\phi^{2}f\wedge\left(\mathcal{\mathfrak{\mathcal{\mathcal{D}}}}A-\bar{\psi}\wedge\gamma_{5}\xi\right)\nonumber \\
 & = & \frac{1}{2\kappa}\int-\frac{1}{2}\phi^{2}\left(\varepsilon_{abcd}R^{ab}\wedge e^{c}\wedge e^{d}-8\bar{\psi}\wedge e_{a}\wedge\gamma^{a}\gamma_{5}\mathcal{\mathfrak{\mathcal{\mathcal{D}}}}\psi\right)\nonumber \\
 &  & +\mathcal{\mathfrak{\mathcal{\mathcal{D}}}}\left(\phi^{2}\varepsilon_{abcd}R^{ab}\wedge B^{cd}-2\phi^{2}f\wedge A\right)-\mathcal{\mathfrak{\mathcal{\mathcal{D}}}}\phi^{2}\left(\varepsilon_{abcd}R^{ab}\wedge B^{cd}-2f\wedge A\right)\nonumber \\
 &  & -\frac{1}{2}\phi^{2}\varepsilon_{abcd}R^{ab}\wedge\left(\bar{\xi}\wedge\sigma^{cd}\psi\right)+4\phi^{2}\mathcal{\mathfrak{\mathcal{\mathcal{D}}}}\bar{\xi}\wedge\gamma_{5}\Psi+2\phi^{2}f\wedge\left(\bar{\psi}\wedge\gamma_{5}\xi\right),\label{eq:L_expand_1}
\end{eqnarray}
where the term $4\phi^{2}\mathcal{\mathfrak{\mathcal{\mathcal{D}}}}\bar{\xi}\gamma_{5}\wedge\Psi$
can be expanded by using the Bianchi identity $\mathcal{\mathfrak{\mathcal{\mathcal{D}}}}\Psi$
given in Eq.(\ref{eq: bianchi}) together with the relation $\gamma_{5}\sigma_{ab}=\frac{i}{2}\varepsilon_{abcd}\sigma^{cd}$
as follows,

\begin{eqnarray}
4\phi^{2}\mathcal{\mathfrak{\mathcal{\mathcal{D}}}}\bar{\xi}\wedge\gamma_{5}\Psi & = & \mathcal{\mathcal{D}}\left(4\phi^{2}\bar{\xi}\wedge\gamma_{5}\Psi\right)+4\mathcal{\mathfrak{\mathcal{\mathcal{D}}}}\phi^{2}\left(\bar{\xi}\wedge\gamma_{5}\Psi\right)\nonumber \\
 &  & +\frac{1}{2}\phi^{2}\varepsilon_{abcd}R^{ab}\wedge\left(\bar{\xi}\wedge\sigma^{cd}\psi\right)-2\phi^{2}f\wedge\left(\bar{\psi}\wedge\delta_{5}\xi\right).\label{eq: term3 lagrangian}
\end{eqnarray}
Substituting Eq.(\ref{eq: term3 lagrangian}) in Eq.(\ref{eq:L_expand_1}),
the free lagrangian density is reduced to the following form, 

\begin{eqnarray}
S_{f} & = & \frac{1}{2\kappa}\int-\frac{1}{2}\phi^{2}\left(\varepsilon_{abcd}R^{ab}\wedge e^{c}\wedge e^{d}-8\bar{\psi}\wedge e_{a}\wedge\gamma^{a}\gamma_{5}\mathcal{\mathfrak{\mathcal{\mathcal{D}}}}\psi\right)\nonumber \\
 &  & -\mathcal{\mathfrak{\mathcal{\mathcal{D}}}}\phi^{2}\left(\varepsilon_{abcd}R^{ab}\wedge B^{cd}-4\bar{\xi}\wedge\gamma_{5}\Psi-2f\wedge A\right)\nonumber \\
 &  & +\mathcal{\mathcal{D}}\left(\phi^{2}\varepsilon_{abcd}R^{ab}\wedge B^{cd}+4\phi^{2}\bar{\xi}\wedge\gamma_{5}\Psi-2\phi^{2}f\wedge A\right),\label{eq: S free}
\end{eqnarray}
where the first part of the resulting action represents the Rarita-Schwinger
term including the compensating scalar field, the second is an extra
term and the third one is total derivative. It can be easily found
that the total action in Eq.(\ref{eq:lagrangian_euler}) is clearly
invariant under the local Lorentz and scale gauge transformations.
Let us now check the invariance of the action in Eq.(\ref{eq: S free})
under the local supersymmetry transformation. To do this, take the
first term in Eq.(\ref{eq: S free}) and name it as $S_{f}^{1}$.
Using Eq.(\ref{eq: var gauge f}), the variation of $S_{f}^{1}$ becomes,

\begin{eqnarray}
\delta_{susy}S_{f}^{1} & = & -\frac{1}{4\kappa}\int-4\phi^{2}i\epsilon_{abcd}R^{ab}\wedge\bar{\varepsilon}\gamma^{c}\psi\wedge e^{d}-8\phi^{2}\mathcal{D}\bar{\varepsilon}\wedge\gamma_{5}\gamma_{a}\mathcal{\mathcal{D}}\psi\wedge e^{a}\nonumber \\
 &  & -8\phi^{2}\bar{\psi}\wedge\gamma_{5}\gamma_{a}\mathcal{\mathcal{D}}^{2}\varepsilon\wedge e^{a}-16i\phi^{2}\bar{\psi}\wedge\gamma_{5}\gamma_{a}\mathcal{\mathcal{D}}\psi\wedge\bar{\varepsilon}\gamma^{a}\psi.\label{eq: var Sf}
\end{eqnarray}
With the help of partial integration and the super-torsion $F^{a}$,
the second term in Eq.(\ref{eq: var Sf}) turns into,

\begin{eqnarray}
-8\phi^{2}\mathcal{\mathcal{D}}\bar{\varepsilon}\wedge\gamma_{5}\gamma_{a}\mathcal{\mathcal{D}}\psi\wedge e^{a} & = & \mathcal{\mathcal{D}}\left(8\phi^{2}\bar{\varepsilon}\gamma_{5}\gamma_{a}\mathcal{\mathcal{D}}\psi\wedge e^{a}\right)+8\mathcal{\mathcal{D}}\phi^{2}\wedge\bar{\varepsilon}\gamma_{5}\gamma_{a}\mathcal{\mathcal{D}}\psi\wedge e^{a}+8\phi^{2}\bar{\varepsilon}\gamma_{5}\gamma_{a}\mathcal{\mathcal{D}}^{2}\psi\wedge e^{a}\nonumber \\
 &  & +8\phi^{2}\bar{\varepsilon}\gamma_{5}\gamma_{a}\mathcal{\mathcal{D}}\psi\wedge F^{a}+8\phi^{2}i\bar{\varepsilon}\gamma_{5}\gamma_{a}\mathcal{\mathcal{D}}\psi\wedge\bar{\psi}\wedge\gamma^{a}\psi.
\end{eqnarray}
Ignoring the total derivative and substituting the last result into
Eq.(\ref{eq: var Sf}), we get,
\begin{eqnarray}
\delta_{susy}S_{f}^{1} & = & -\frac{1}{4\kappa}\int-4i\phi^{2}\epsilon_{abcd}R^{ab}\wedge\bar{\varepsilon}\gamma^{c}\psi\wedge e^{d}+8\phi^{2}\bar{\varepsilon}\wedge\gamma_{5}\gamma_{a}\mathcal{\mathcal{D}}^{2}\psi\wedge e^{a}\nonumber \\
 &  & +8\phi^{2}\bar{\varepsilon}\wedge\gamma_{5}\gamma_{a}\mathcal{\mathcal{D}}\psi\wedge F^{a}+8i\phi^{2}\bar{\varepsilon}\wedge\gamma_{5}\gamma_{a}\mathcal{\mathcal{D}}\psi\wedge\bar{\psi}\wedge\gamma^{a}\psi\nonumber \\
 &  & -8\phi^{2}\bar{\psi}\wedge\gamma_{5}\gamma_{a}\mathcal{\mathcal{D}}^{2}\varepsilon\wedge e^{a}-16\phi^{2}\bar{\psi}\wedge\gamma_{5}\gamma_{a}\mathcal{\mathcal{D}}\psi\wedge\bar{\varepsilon}\gamma^{a}\psi\nonumber \\
 &  & +8\mathcal{\mathcal{D}}\phi^{2}\wedge\bar{\varepsilon}\wedge\gamma_{5}\gamma_{a}\mathcal{\mathcal{D}}\psi\wedge e^{a}.\label{eq: var Sf 2}
\end{eqnarray}
Considering the second and fifth terms in Eq.(\ref{eq: var Sf 2}),
one obtains,

\begin{eqnarray}
8\phi^{2}\bar{\varepsilon}\gamma_{5}\gamma_{a}\mathcal{\mathcal{D}}^{2}\psi\wedge e^{a}-8\phi^{2}\bar{\psi}\wedge\gamma_{5}\gamma_{a}\mathcal{\mathcal{D}}^{2}\varepsilon\wedge e^{a} & = & 4i\phi^{2}\epsilon_{abcd}R^{ab}\wedge\bar{\varepsilon}\gamma^{c}\psi\wedge e^{d}.
\end{eqnarray}
Here, $\gamma_{5}\gamma_{a}\gamma_{bc}=i\gamma_{5}\eta_{a[b}\gamma_{c]}+\epsilon_{abcd}\gamma^{d}$
and $\mathcal{D}^{2}\varUpsilon=-\frac{i}{4}R^{ab}\gamma_{ab}\varUpsilon+w\left(\varUpsilon\right)f\varUpsilon$
have been used where $\varUpsilon$ is a spinor field. Therefore,
the last result and the first term in Eq.(\ref{eq: var Sf 2}) cancel
each other, so we get,

\begin{eqnarray}
\delta_{susy}S_{f}^{1} & = & -\frac{1}{4\kappa}\int8\bar{\varepsilon}\gamma_{5}\gamma_{a}\mathcal{\mathcal{D}}\psi\wedge F^{a}+8\mathcal{\mathcal{D}}\phi^{2}\wedge\bar{\varepsilon}\gamma_{5}\gamma_{a}\mathcal{\mathcal{D}}\psi\wedge e^{a}\nonumber \\
 &  & +8i\phi^{2}\left(\bar{\varepsilon}\gamma_{5}\gamma_{a}\mathcal{\mathcal{D}}\psi\wedge\bar{\psi}\wedge\gamma^{a}\psi+2\bar{\varepsilon}\gamma^{a}\psi\wedge\bar{\psi}\wedge\gamma_{5}\gamma_{a}\mathcal{\mathcal{D}}\psi\right).\label{eq: var Sf 3}
\end{eqnarray}
Using the relations $\psi\wedge\bar{\psi}=\frac{1}{2}\left(\bar{\psi}\wedge\gamma^{a}\psi\right)\gamma_{a}-\frac{1}{8}\left(\bar{\psi}\wedge\gamma^{ab}\psi\right)\gamma_{ab}$,
$\gamma_{a}\gamma^{bc}\gamma^{a}=0$ and $\gamma^{a}\gamma^{b}\gamma_{5}\gamma_{a}=2\gamma^{b}\gamma_{5}$,
the fourth term becomes,

\begin{equation}
2\bar{\varepsilon}\gamma^{a}\psi\wedge\bar{\psi}\wedge\gamma_{5}\gamma_{a}\mathcal{\mathcal{D}}\psi=-\bar{\varepsilon}\gamma_{5}\gamma_{a}\mathcal{\mathcal{D}}\psi\wedge\left(\bar{\psi}\wedge\gamma^{a}\psi\right),
\end{equation}
then substituting the last equation into Eq.(\ref{eq: var Sf 3}),
the variation reduces to
\begin{eqnarray}
\delta_{susy}S_{f}^{1} & = & -\frac{1}{4\kappa}\int8\bar{\varepsilon}\gamma_{5}\gamma_{a}\mathcal{\mathcal{D}}\psi\wedge F^{a}-8\mathcal{D}\phi^{2}\wedge e_{a}\wedge\bar{\varepsilon}\gamma^{a}\gamma_{5}\mathcal{D}\psi.\label{eq: var Sf 4}
\end{eqnarray}
Assuming the constraints as $F^{a}=0$ and $\mathcal{\mathfrak{\mathcal{\mathcal{D}}}}\phi=0$,
$\delta_{susy}S_{f}^{1}$ becomes zero. Under these constraints, Eq.(\ref{eq: S free})
finally reduces to pure scale-invariant supergravity action with a
boundary term,

\begin{eqnarray}
S_{f} & = & \frac{1}{2\kappa}\int-\frac{1}{2}\phi^{2}\left(\varepsilon_{abcd}R^{ab}\wedge e^{c}\wedge e^{d}-8\bar{\psi}\wedge e_{a}\wedge\gamma^{a}\gamma_{5}\mathcal{\mathfrak{\mathcal{\mathcal{D}}}}\psi\right)\nonumber \\
 &  & +\mathcal{\mathcal{D}}\left(\phi^{2}\varepsilon_{abcd}R^{ab}\wedge B^{cd}+4\phi^{2}\bar{\xi}\wedge\gamma_{5}\Psi-2\phi^{2}f\wedge A\right).
\end{eqnarray}
Therefore, the action $S_{f}$ leave invariant under the local supersymmetry
transformations in terms of given conditions. The additional fields
$B^{ab}\left(x\right)$, $\xi^{\alpha}\left(x\right)$ and $A\left(x\right)$
appeared only in the boundary term so we can say that they do not
affect the dynamics of the action. This result also shows that $s\mathcal{MW}$
algebra provides an alternative way to construct the supergravity
lagrangian including the local scale invariance. Finally, ignoring
the boundary term and applying the condition to $L_{0}$, the total
action in Eq.(\ref{eq:lagrangian_euler}) takes the following form,
\begin{eqnarray}
S & = & \int-\frac{1}{4\kappa}\phi^{2}\left(\varepsilon_{abcd}R^{ab}\wedge e^{c}\wedge e^{d}-8\bar{\psi}\wedge e_{a}\wedge\gamma^{a}\gamma_{5}\mathcal{\mathfrak{\mathcal{\mathcal{D}}}}\psi\right)+\frac{1}{4}f\wedge\,^{*}f+\frac{\lambda}{4}\phi^{4}\,^{*}1.\label{eq:lagrangian_euler-1}
\end{eqnarray}

The equations of motion can be found by taking variations of Eq.(\ref{eq:lagrangian_euler-1})
with respect to gauge fields $\omega^{ab}\left(x\right)$, $e^{a}\left(x\right)$,
$\chi\left(x\right)$, $\bar{\psi}^{\alpha}\left(x\right)$ and $\phi\left(x\right)$,
respectively,

\begin{eqnarray}
0 & = & \varepsilon_{abcd}\mathcal{\mathfrak{\mathcal{\mathcal{D}}}}e^{c}\wedge e^{d}+i\bar{\psi}\wedge e^{c}\wedge\gamma_{c}\sigma_{ab}\gamma_{5}\psi,\label{eq: var omega}\\
0 & = & \frac{\phi^{2}}{\kappa}\left(\varepsilon_{abcd}R^{bc}\wedge e^{d}+4\bar{\psi}\wedge\gamma_{a}\gamma_{5}\mathcal{\mathfrak{\mathcal{\mathcal{D}}}}\psi\right)\nonumber \\
 &  & +\frac{1}{2}\left(f_{ab}e^{b}\wedge\,^{*}f-\frac{1}{2}\varepsilon_{abcd}f^{bc}e^{d}\wedge f\right)-\frac{\lambda\phi^{4}}{2}\,^{*}e_{a},\label{eq:var_S_e}\\
0 & = & \frac{\phi^{2}}{\kappa}\bar{\psi}\wedge e_{a}\wedge\gamma^{a}\gamma_{5}\psi+\frac{1}{2}\mathcal{\mathfrak{\mathcal{\mathcal{D}}}}\,^{*}f,\\
0 & = & e^{c}\wedge\gamma_{c}\gamma_{5}\mathcal{\mathfrak{\mathcal{\mathcal{D}}}}\psi,\\
0 & = & \left(\varepsilon_{abcd}R^{ab}\wedge e^{c}\wedge e^{d}-8\bar{\psi}\wedge e_{a}\wedge\gamma^{a}\gamma_{5}\mathcal{\mathfrak{\mathcal{\mathcal{D}}}}\psi\right)-2\kappa\lambda\phi^{2}\,^{*}1.
\end{eqnarray}
Let us analyze Eq.(\ref{eq: var omega}) which is known as the torsion
equation, 
\begin{eqnarray}
\varepsilon_{abcd}\mathcal{D}e^{c}\wedge e^{d} & = & -i\bar{\psi}\wedge e^{c}\gamma_{c}\sigma_{ab}\gamma_{5}\wedge\psi\nonumber \\
 & = & -\frac{1}{2}\bar{\psi}\wedge\gamma_{5}e^{c}\gamma_{c}\left(\gamma_{a}\gamma_{b}-\gamma_{b}\gamma_{a}\right)\wedge\psi\nonumber \\
 & = & i\epsilon_{abcd}\bar{\psi}\gamma^{c}\wedge\psi\wedge e^{d},\label{eq: zero st}
\end{eqnarray}
where we used the expressions $\gamma_{c}\gamma_{a}\gamma_{b}=\eta_{ca}\gamma_{b}-\eta_{cb}\gamma_{a}+\eta_{ab}\gamma_{c}-i\epsilon_{cabd}\gamma_{5}\gamma^{d}$
and $\bar{\psi}\gamma_{5}\gamma_{a}\psi=0.$ If we compare the both
sides of Eq.(\ref{eq: zero st}), we obtain $\mathcal{D}e^{c}=i\left(\bar{\psi}\gamma^{c}\wedge\psi\right)$
which satisfies the super-torsion constraint ($F^{a}=0$). Moreover,
using Eq.(\ref{eq:var_S_e}) and transforming the tangent indices
to world space-time indices, one can show that the generalized Einstein
field equations can be written as follows,

\begin{equation}
R_{\,\,\alpha}^{\mu}-\frac{1}{2}\delta_{\alpha}^{\mu}R=T\left(\psi\right)_{\,\,\alpha}^{\mu}-\kappa\phi^{-2}\left\{ T\left(\phi\right)_{\,\,\alpha}^{\mu}+\frac{1}{2}T\left(f\right)_{\,\,\alpha}^{\mu}\right\} ,\label{eq: einstein_filed_eq}
\end{equation}
where,

\begin{eqnarray}
T\left(\psi\right)_{\,\,\alpha}^{\mu} & = & \bar{\psi_{\nu}}\gamma_{\alpha}\gamma_{5}\mathcal{\mathfrak{\mathcal{\mathcal{D}}}}_{[\rho}\psi_{\sigma]}\varepsilon^{\mu\nu\rho\sigma},\nonumber \\
T\left(\phi\right)_{\,\,\alpha}^{\mu} & = & \frac{\lambda}{4}\delta_{\alpha}^{\mu}\phi^{4},\nonumber \\
T\left(f\right)_{\,\,\alpha}^{\mu} & = & f_{\beta\alpha}f^{\mu\beta}+\frac{1}{4}\delta_{\alpha}^{\mu}f^{\gamma\delta}f_{\gamma\delta}.
\end{eqnarray}
Here, as we mention before, we want to note that the additional spinorial
field $\xi^{\alpha}\left(x\right)$ does not a new gravitino field
because it contributes to neither the spacetime coordinate nor to
the equations of motion. It only contributes to the boundary term.

Let us analyze the constraint $\mathcal{D}\phi=d\phi-\chi\phi=0$
which was proposed above to keep the scale invariance of the total
action. This constraint gives a relation between the scalar compensating
field and the dilatation gauge field as,

\begin{equation}
\chi\left(x\right)=d\ln\phi\left(x\right).\label{eq: condition ksi}
\end{equation}
Taking the exterior differential of the last equation, we find the
dilatation curvature to be zero (i.e., $f=d\chi=0$). This is another
way of expressing the inverse Higgs effect \citep{ivanov1975,low2002}.
Clearly, this also shows the equivalence of the inverse Higgs constraint
to equations of motion. From this constraint, the dilatation gauge
field can be alternatively written as $\chi\left(x\right)=d\alpha\left(x\right)$,
where $\alpha\left(x\right)$ is a coordinate dependent field. Therefore,
the scalar field can be described as a function of $\alpha\left(x\right)$,
\begin{equation}
\phi\left(x\right)=e^{\alpha\left(x\right)}\label{eq: confition phi}
\end{equation}
Since $f=0$, the expression $T\left(f\right)_{\,\,\alpha}^{\mu}$
goes to zero. As a result, Eq.(\ref{eq: einstein_filed_eq}) reduces
to,

\begin{equation}
R_{\,\,\alpha}^{\mu}-\frac{1}{2}\delta_{\alpha}^{\mu}R+\Lambda\delta_{\alpha}^{\mu}=T\left(\psi\right)_{\,\,\alpha}^{\mu},\label{eq: einstein f e cosm}
\end{equation}
where the cosmological term emerged as a function of the field $\alpha\left(x\right)$
as follows,
\begin{equation}
\Lambda\left(x\right)=\frac{\kappa\lambda}{4}e^{2\alpha\left(x\right)}\label{eq: cosm term}
\end{equation}

\section{Conclusion}

In this paper, we proposed an alternative way to obtain $D=4$, $\mathcal{N}=1$
supergravity lagrangian including scale transformations by using the
Maxwell-Weyl superalgebra. This superalgebra contains two fermionic
generators $Q_{\alpha}$ and $\Sigma_{\alpha}$, where the first one
is the standard $\mathcal{N}=1$ fermionic generator and the other
is required for the supersymmetrization of the Maxwell generator $Z_{ab}$
\citep{kamimura2012}. The new additional fermionic generators $\Sigma_{\alpha}$
were originally introduced by Green \citep{green1989} in the context
of superstring theory (see also \citep{bergshoeff1995}). Therefore,
the Maxwell superalgebras can be considered as an extension of the
Green algebra by adding the tensorial charges $Z_{ab}$.

We constructed the gauge theory of gravity based on $s\mathcal{MW}$
algebra. In this framework, we obtained the transformation of gauge
fields, curvatures and the Bianchi identities of $s\mathcal{MW}$
algebra. The corresponding four-form lagrangian was constructed with
bilinear of the curvatures together with a compensating scalar field.
Our result can be considered as the Weyl enlargement of the paper
\citep{azcarraga2014} or/and supersymmetric generalization of the
Maxwell-Weyl gravity \citep{cebecioglu2014}. According to Eq.(\ref{eq: S free}),
the Maxwell fields $B^{ab}\left(x\right)$, the new fermionic fields
$\xi^{\alpha}\left(x\right)$ and the gauge field $A\left(x\right)$
appeared only in the boundary term. This result shows that the mentioned
fields do not have a dynamical character, but the corresponding curvatures
play an essential role for constructing an invariant action. We can
say that analyzing the boundary term to find a physical or geometrical
properties of the related fields is an open problem \citep{regge1974,hawking1996,easson2011}.
After analyzing the condition $\mathcal{D}\phi=0$, we found the curvature
of dilatation field goes to zero $f=d\chi=0$. This condition also
represents the Maurer-Cartan form of the dilatation field. Therefore,
we have obtained an alternative way to express the inverse Higgs effect
\citep{ivanov1975,low2002}. Moreover, we obtained a relation between
the cosmological constant and the dilatation gauge field in Eq.(\ref{eq: cosm term}). 

We remark that the physical interpretation of the Maxwell symmetry
and its supersymmetric extension are under research. According to
the early studies \citep{bacry1970,schrader1972}, the Maxwell symmetry
has been used to describe a particle moving in a Minkowski spacetime
filled with a constant EM background field, parametrized by additional
degrees of freedom related to the central charges $Z_{ab}$. So this
symmetry was considered as the symmetry group of a particle moving
in a constant EM field \citep{bonanos2010B}. Later, the Maxwell algebra
was used as an alternative way to generalize Einstein's theory of
gravity and supergravity. In the gravitational framework, the additional
degrees of freedom represent uniform gauge field strengths in (super)space
which lead to uniform constant energy density \citep{bonanos2010}.
By gauging the Maxwell (super)algebra, a generalized cosmological
constant was proposed in \citep{azcarraga2011,concha2015A,durka2012}.
In addition, the Maxwell gauge fields $B_{ab}\left(x\right)$ contributed
to the stress energy-momentum tensor $T_{ab}\left(B\right)$ \citep{azcarraga2011,soroka2012,durka2011,cebecioglu2014,concha2015A,durka2012,penafiel2018}.
To our knowledge, this energy-momentum term has not been analyzed
yet, but it is known that such an additional term may be related to
the dark energy \citep{frieman2008,padmanabhan2009}. Also, the Maxwell
symmetry provides a geometric background to define vector inflatons
in cosmological models \citep{azcarraga2013A}. This symmetry was
used to describe higher spin fields \citep{fedoruk2013,fedoruk2013A},
planar dynamics of the Landau problem \citep{fedoruk2012} and it
was also applied to the string theory as an internal symmetry of the
matter gauge fields \citep{hoseinzadeh2015}. Moreover, the cosmological
consequences of the Maxwell symmetry are still an open problem. So,
these results show the importance of the Maxwell symmetry. 

Finally, considering the results in \citep{azcarraga2014,concha2015A,durka2012,concha2014,concha2015B}
plus our findings, it is shown that the Maxwell superalgebras provide
an effective geometric framework to construct supergravity lagrangians
by using its curvatures. 
\begin{acknowledgments}
The work of S.K. has been partly supported by the Scientific and Technological
Research Council of Turkey (TUB\.{I}TAK) under the grant number 2214-A.
The authors also wish to thank J. Kowalski-Glikman for useful discussions
and their kind hospitality at Institute for Theoretical Physics, University
of Wroc\l aw, where a part of this work was done.
\end{acknowledgments}

\section{NOTATIONS AND CONVENTIONS}

Here, we summarize our notation and conventions in $D=4$. The Dirac
gamma matrices are defined by the well-known expression $\left\{ \gamma^{a},\gamma^{b}\right\} =2\eta^{ab}$
where $\eta_{ab}$ is the (mostly minus) Minkowski metric and these
matrices obey the following relations,

\begin{equation}
\gamma^{5}=\gamma_{5}=i\gamma^{0}\gamma^{1}\gamma^{2}\gamma^{3},\,\,\,\,\,\,\,\left(\gamma^{5}\right)^{2}=1,\,\,\,\,\,\,\,\,\left(\gamma_{5}\right)^{\dagger}=\gamma_{5},
\end{equation}

\begin{equation}
\sigma^{ab}=\frac{i}{2}\left[\gamma^{a},\gamma^{b}\right],\,\,\,\,\,\,\,\,\,\gamma_{5}\sigma_{ab}=\frac{i}{2}\varepsilon_{abcd}\sigma^{cd},
\end{equation}

\begin{equation}
\left\{ \gamma^{a},\gamma^{5}\right\} =0,\,\,\,\,\,\,\,\,\,\left[\sigma^{ab},\gamma^{5}\right]=0,
\end{equation}

\begin{eqnarray}
\left[\sigma^{ab},\gamma^{c}\right] & = & 2i\left(\eta^{[b|c}\gamma^{a]}\right),\,\,\,\,\,\,\,\,\,\,\left\{ \sigma^{ab},\gamma^{c}\right\} =2\varepsilon^{abcd}\gamma_{5}\gamma_{d},\\
\left[\sigma^{ab},\gamma^{5}\gamma^{c}\right] & = & 2i\gamma^{5}\left(\eta^{[b|c}\gamma^{a]}\right),\,\,\,\,\,\,\,\,\,\,\left\{ \sigma^{ab},\gamma^{5}\gamma^{c}\right\} =2\varepsilon^{abcd}\gamma_{d},\\
\left[\sigma^{ab},\sigma^{cd}\right] & = & i\left(\eta^{a[d}\sigma^{b|c]}+\eta^{b[c}\sigma^{a|d]}\right),\\
\left\{ \sigma^{ab},\sigma^{cd}\right\}  & = & 2\left(\eta^{a[c}\eta^{b|d]}+i\varepsilon^{abcd}\gamma^{5}\right),
\end{eqnarray}
where $\sigma^{ab}=-\sigma^{ba}$ are $O\left(3,1\right)$ Lorentz
generators and the antisymmetrization is defined by $A^{[c}B^{d]}=A^{c}B^{d}-A^{d}B^{c}$.
Moreover, we have,

\begin{equation}
\gamma_{a}\sigma^{ab}=3i\gamma^{b},\,\,\,\,\,\,\sigma^{ab}\sigma_{ab}=12,
\end{equation}

\begin{equation}
\gamma_{c}\sigma^{ab}\gamma^{c}=0,\,\,\,\,\,\,\,\sigma_{ab}\gamma_{c}\sigma^{ab}=0,
\end{equation}

\begin{equation}
\sigma_{ab}\sigma_{cd}\sigma^{ab}=0,\,\,\,\,\,\,\,\varepsilon_{abcd}\sigma^{cd}=-2i\gamma^{5}\sigma_{ab}.
\end{equation}

In this work, we use the Majorana spinors which satisfy $\bar{\psi}=\psi^{T}C$.
Here, $C=\gamma_{0}$ represents the charge conjugation matrix and
satisfies the following relations,

\begin{equation}
C^{T}=C^{-1}=-C,\,\,\,\,\,\,\,C^{2}=-1,
\end{equation}

\begin{equation}
\left(C\gamma_{a}\right)^{T}=\left(C\gamma_{a}\right),\,\,\,\,\,\,\,\left(C\sigma_{ab}\right)^{T}=\left(C\sigma_{ab}\right),
\end{equation}

\begin{equation}
\left(C\gamma_{5}\right)^{T}=-\left(C\gamma_{5}\right),\,\,\,\,\,\,\,\left(C\gamma_{a}\gamma_{5}\right)^{T}=-\left(C\gamma_{a}\gamma_{5}\right),
\end{equation}
where $T$ denotes the transpose of the corresponding matrix. For
the closure of the superalgebras, the following cyclic identities
are required,

\begin{equation}
\left(C\gamma_{e}\right)_{(\alpha\beta}\left(C\gamma^{e}\right)_{\rho\sigma)}=0.
\end{equation}

The spinor variables satisfy the following relations,

\begin{eqnarray}
\psi_{\alpha}\chi_{\beta} & = & -\chi_{\beta}\psi_{\alpha},\,\,\,\,\,\,\,\psi\chi=\chi\psi,\\
\bar{\psi}\chi & = & \bar{\chi}\psi,\,\,\,\,\,\,\,\bar{\psi}\bar{\chi}=\bar{\chi}\bar{\psi},\\
\bar{\psi}\gamma_{a}\chi & = & -\bar{\chi}\gamma_{a}\psi,\,\,\,\,\,\,\,\bar{\psi}\gamma_{a}\psi=0,\\
\bar{\psi}\sigma_{ab}\chi & = & -\bar{\chi}\sigma_{ab}\psi,\,\,\,\,\,\,\,\bar{\psi}\sigma_{ab}\psi=0,\\
\bar{\psi}\gamma_{5}\gamma_{a}\chi & = & \bar{\chi}\gamma_{5}\gamma_{a}\psi,\,\,\,\,\,\,\,\bar{\psi}\gamma_{5}\chi=\bar{\chi}\gamma_{5}\psi.
\end{eqnarray}
If spinors have differential form structure, we have to consider the
following relationships,

\begin{eqnarray}
\psi_{\alpha}^{\left(p\right)}\wedge\chi_{\beta}^{\left(q\right)} & = & -\left(-1\right)^{p\cdot q}\chi_{\beta}\wedge\psi_{\alpha},\\
\bar{\psi}^{\left(p\right)}\wedge\chi{}^{\left(q\right)} & = & \left(-1\right)^{p\cdot q}\bar{\chi}\wedge\psi.
\end{eqnarray}

Considering a one-form spinor $\psi$, we have the following Fierz
identities, 

\begin{eqnarray}
\psi\wedge\bar{\psi} & = & \frac{1}{2}\gamma_{a}\bar{\psi}\wedge\gamma^{a}\psi-\frac{1}{8}\sigma_{ab}\bar{\psi}\wedge\sigma^{ab}\psi,\\
\gamma_{a}\psi\wedge\bar{\psi}\wedge\gamma^{a}\psi & = & \sigma_{ab}\psi\wedge\bar{\psi}\wedge\sigma^{ab}\psi=0.
\end{eqnarray}


\begin{thebibliography}{10}
\bibitem{bacry1970} H. Bacry, P. Combe, J. L. Richard, Nuovo Cimento
\textbf{67}, 267-299 (1970).

\bibitem{schrader1972} R. Schrader, Fortschritte der Physik \textbf{20},
701 (1972).

\bibitem{soroka2005} D. V. Soroka, V. A. Soroka, Phys. Lett. B \textbf{607},
302-305 (2005).

\bibitem{azcarraga2011} J. A. de Azcárraga, K. Kamimura, J. Lukierski,
Phys. Rev. D \textbf{83}, 124036, (2011). 

\bibitem{soroka2012} D. V. Soroka, V. A. Soroka, Phys. Lett. B \textbf{707},
160-162 (2012). 

\bibitem{durka2011} R. Durka, J. Kowalski-Glikman, M. Szczachor,
Mod. Phys. Lett. A \textbf{26}, 2689 (2011).

\bibitem{cebecioglu2014} O. Cebecio\u{g}lu, S. Kibaro\u{g}lu, Phys.
Rev. D \textbf{90}, 084053 (2014).

\bibitem{frieman2008} J. Frieman, M. Turner, D. Huterer, Ann. Rev.
Astron. Astrophys. \textbf{46}, 385 (2008).

\bibitem{padmanabhan2009} T. Padmanabhan, Adv. Sci. Lett. \textbf{2},
174 (2009).

\bibitem{bonanos2010} S. Bonanos, J. Gomis, K. Kamimura, J. Lukierski,
Phys. Rev. Lett. \textbf{104}, 090401 (2010).

\bibitem{dAuria1982} R. D\textquoteright Auria, P. Fré, Nucl. Phys.
B \textbf{201,} 101 (1982).

\bibitem{green1989} M. B. Green, Phys. Lett. B \textbf{223}, 157\textendash 164
(1989).

\bibitem{azcarraga2013} J. A. de Azcárraga, J. M. Izquierdo, J. Lukierski,
M. Woronowicz, Nucl. Phys. B \textbf{869}, 303 (2013).

\bibitem{concha2014A} P. K. Concha, E. K. Rodríguez, Nucl. Phys.
B \textbf{886}, 1128 (2014).

\bibitem{azcarraga2004} J. A. de Azcárraga, J. M. Izquierdo, M. Picon,
O. Varela, Class. Quantum Gravity \textbf{21}, S1375 (2004).

\bibitem{izaurieta2006} F. Izaurieta, E. Rodríguez, P. Salgado, J.
Math. Phys. \textbf{47}, 123512 (2006).

\bibitem{azcarraga2014} J. A. de Azcárraga, J. M. Izquierdo, Nucl.
Phys. B \textbf{885}, 34-45 (2014).

\bibitem{concha2015A} P. K. Concha, E. K. Rodriguez, P. Salgado,
JHEP \textbf{08}, 009 (2015). 

\bibitem{durka2012} R. Durka, Kowalski-Glikman, M. Szczachor, Mod.
Phys. Lett. A \textbf{27}, 1250023 (2012).

\bibitem{lukierski2011} J. Lukierski, Proc. Steklov Inst. Math. \textbf{272},
1-8 (2011).

\bibitem{kamimura2012} K. Kamimura, J. Lukierski, Phys. Lett. B \textbf{707},
292-297 (2012).

\bibitem{gomis2009} J. Gomis, K. Kamimura, J. Lukierski, JHEP \textbf{08},
39 (2009).

\bibitem{bonanos2009} S. Bonanos, J. Gomis, J. Phys. A: Math. Theor.
\textbf{42}, 145206 (2009).

\bibitem{bonanos2010A} S. Bonanos, J. Gomis, K. Kamimura, J. Lukierski,
J. Math. Phys. \textbf{51}, 102301 (2010).

\bibitem{fedoruk2013} S. Fedoruk, J. Lukierski, JHEP \textbf{02},
128 (2013).

\bibitem{fedoruk2013A} S. Fedoruk, J. Lukierski, J. Phys. Conf. Ser.
\textbf{474}, 012016 (2013).

\bibitem{fedoruk2012} S. Fedoruk, J. Lukierski, Phys. Lett. B \textbf{718},
646 (2012).

\bibitem{hoseinzadeh2014} S. Hoseinzadeh, A. Rezaei-Aghdam, Phys.
Rev. D \textbf{90}, 084008 (2014).

\bibitem{concha2014} P. K. Concha, E. K. Rodriguez, JHEP \textbf{09},
090 (2014). 

\bibitem{concha2015B} P. K. Concha, O. Fierro, E. K. Rodriguez, P.
Salgado, Phys. Lett. B \textbf{750}, 117-121 (2015). 

\bibitem{cebecioglu2015} O. Cebecio\u{g}lu, S. Kibaro\u{g}lu, Phys.
Lett. B \textbf{751}, 131-134 (2015).

\bibitem{kibaroglu2018} S. Kibaro\u{g}lu, O. Cebecio\u{g}lu, Mod.
Phys. Lett. A \textbf{34}, 1950016 (2019).

\bibitem{penafiel2017} D. M. Peñafiel, L. Ravera, Fortschr. Phys.
\textbf{65}, 1700005 (2017).

\bibitem{penafiel2018} D. M. Peñafiel, L. Ravera, Eur. Phys. J. C
\textbf{78}, 945 (2018).

\bibitem{ravera2018} L. Ravera, Eur. Phys. J. C \textbf{78}, 211
(2018).

\bibitem{omote1971} M. Omote, Lett. Nuovo Cimento \textbf{2}, 58
(1971).

\bibitem{bregman1973} A. Bregman, Prog. Theor. Phys. \textbf{49},
667 (1973).

\bibitem{charap1974} J. M. Charap, W. Tait, Proc. R. Soc. A \textbf{340},
249 (1974).

\bibitem{kasuya1975} M. Kasuya, Nuovo Cimento Soc. Ital. Fis. \textbf{28B},
127 (1975).

\bibitem{babourova2006} O. V. Babourova, B. N. Frolov, V. Ch. Zhukovsky,
Phys. Rev. D \textbf{74}, 064012 (2006).

\bibitem{blagojevic2002} M. Blagojevic, \textit{Gravitation and Gauge
Symmetries} (IOP, Bristol, 2002).

\bibitem{weyl1961} H. Weyl, \textit{Space, Time, Matter} (Dover,
New York, 1961). 

\bibitem{brans1961} C. Brans, R. H. Dicke, Phys. Rev. \textbf{124},
925 (1961). 

\bibitem{dirac1973} P. A. M. Dirac, Proc. R. Soc. A \textbf{333},
403 (1973). 

\bibitem{dereli1982} T. Dereli, R.W. Tucker, Phys. Lett. \textbf{110B},
206 (1982). 

\bibitem{agnese1975} A. G. Agnese, Phys. Rev. D \textbf{12}, 3804
(1975).

\bibitem{fradkin1985} E. S. Fradkin, A. A. Tseytlin, Phys. Rep. \textbf{119},
233-362 (1985).

\bibitem{kaku1978} M. Kaku, P. K. Townsend, P. van Nieuwenhuizen,
Phys. Rev. D \textbf{17}, 3179 (1978).

\bibitem{nieuwenhuizen1986} P. van Nieuwenhuizen, Int. J. Mod. Phys.
A \textbf{1}, 155-191 (1986).

\bibitem{nieuwenhuizen1981} P. van Nieuwenhuizen, Phys. Rep. \textbf{68},
189-398 (1981).

\bibitem{ivanov1975} E. A. Ivanov, V. I. Ogievetsky, Teor. Mat. Fiz.
\textbf{25}, 164 (1975).

\bibitem{low2002} I. Low, A. V. Manohar, Phys. Rev. Lett. \textbf{88},
101602 (2002).

\bibitem{bergshoeff1995} E. Bergshoeff, E. Sezgin, Phys. Lett. B
\textbf{354}, 256 (1995).

\bibitem{regge1974} T. Regge, C. Teitelboim, Ann. Phys. \textbf{88},
286-318 (1974).

\bibitem{hawking1996} S. W. Hawking, G. T. Horowitz, Class. Quantum
Grav. \textbf{13}, 1487-1498 (1996).

\bibitem{easson2011} D. A. Easson, P. H. Frampton, G. F. Smoot, Phys.
Lett. B 696, 273-277 (2011).

\bibitem{bonanos2010B} S. Bonanos, J. Gomis, J. Phys. A \textbf{43},
015201 (2010).

\bibitem{azcarraga2013A} J. A. de Azcárraga, K. Kamimura, J. Lukierski,
Int. J. Mod. Phys. Conf. Ser. \textbf{23}, 01160 (2013).

\bibitem{hoseinzadeh2015} S. Hoseinzadeh, A. Rezaei-Aghdam, Eur.
Phys. J. C \textbf{75}, 227 (2015).
\end{thebibliography}
\end{document}